\documentclass[acus]{JAC2000}

\usepackage{graphicx}

\begin{document}
\title{Determining Phase-Space Properties of the LEDA RFQ Output
    Beam\thanks{Work supported by US Department of Energy}}

\author{W.P. Lysenko, J.D. Gilpatrick, L.J. Rybarcyk, J.D. Schneider,
        H.V. Smith, Jr., and L.M. Young,\\ LANL, Los Alamos, NM 87545, USA\\ 
        M.E. Schulze, General Atomics, Los Alamos, NM 87544, USA}

\maketitle

\begin{abstract} 
Quadrupole scans were used to characterize the {\sc leda} {\sc rfq}
beam.  Experimental data were fit to computer simulation models for
the rms beam size.  The codes were found to be inadequate in
accurately reproducing details of the wire scanner data.  When this
discrepancy is resolved, we plan to fit using all the data in wire
scanner profiles, not just the rms values, using a 3-D nonlinear code.
\end{abstract}

\section{INTRODUCTION}

During commissioning of the {\sc leda rfq}\cite{r1,r2}, we found
that the beam behaved in the high energy beam transport ({\sc hebt})
much as predicted.  Thus the actual {\sc rfq} beam must have been
close to that computed by the {\sc parmteqm} code.

The {\sc hebt} included only limited diagnostics\cite{r3} but we were
able to get additional information on the {\sc rfq} beam distribution
using quadrupole scans\cite{r4}. An good understanding of the {\sc
rfq} beam and beam behavior in the {\sc hebt} will be helpful for the
upcoming beam halo experiment.  The problems with the quad scan
measurements were the strong space effects and the almost complete
lack of knowledge of the longitudinal phase space.  Also, our
simulation codes, which served as the models for the data fitting, did
not accurately reproduce the measured beam profiles at the wire
scanner.

\section{HEBT DESIGN}

The {\sc hebt}\cite{r5} transports the {\sc rfq} beam to the beamstop
and provides space for beam diagnostics.  Here, we discuss {\sc hebt}
properties relevant to beam characterization.

\begin{Itemize}
\item{\em Design has Weak Focusing.}  Ideally, the {\sc hebt} would
have closely-space quadrupoles at the upstream end until the beam is
significantly debunched, i.e., for about one meter.  After this point,
we could use any kind of matching scheme with no fear of spoiling the
beam distribution with space-charge nonlinearities.

Our {\sc hebt} design uses four quadrupoles, which is the minimum that
provides adequate focusing for the given length.  Any fewer than four
quadrupoles results in the generation of long Gaussian-like tails in
the beam, which would be scraped off in the {\sc hebt}.

\item{\em Good Tune is Important.}  If a tune has a small waist in
the upstream part of the {\sc hebt}, the beam will also acquire
Gaussian-like tails.  Simulations showed that good tunes existed for
our four-quadrupole beamline and were stable (slight changes in magnet
settings or input beam did not lead to beam degradation).

\item{\em Beam Size Control.}  In our design, increasing the strength
of the last quadrupole (Q4) increases the beam size in both $x$
and $y$ by about the same amount.  This is because
there is a crossover in $x$ just downstream of Q4 and a (virtual)
crossover just upstream of Q4 in $y$.  If the beam turns out to not be
circular, this can be adjusted by Q3, which moves the upstream
crossover point.

\item{\em Emittance Growth in HEBT.}  Simulations showed that the
transverse emittances grew by about 30\% in the {\sc hebt}.  However,
this did not affect final beam size.  At the downstream end of the
{\sc hebt} and in the beamstop, the beam is in the zero-emittance
regime (very narrow phase-space ellipses).  Simulations with {\sc
trace 3-d}, which has no nonlinear effects, and a 3-D particle code
that included nonlinear space-charge predicted almost identical final
beam sizes.
\end{Itemize}

\section{OBSERVED HEBT PERFORMANCE}

Near the beamstop entrance, there is a collimator with a size less
than 3 times the rms beam size.  Initial runs showed beam hitting the
top and bottom of the the collimator, indicating the beam was too
large in $y$.  This was fixed by readjusting Q3 and slightly reducing
Q4 to reduce the beam size.  After these
adjustments, beam losses were negligible.  This indicated the {\sc
hebt} was operating as predicted and the {\sc rfq} beam was about as
predicted.  There were no long tails generated in the {\sc hebt} that
were being scraped off.  Thus our somewhat risky design, having only
four quadrupoles, worked as designed.

\section{QUADRUPOLE SCANS}

\subsection{Procedure}

Only the first two quadrupoles were used.  For characterizing the beam
in $y$, Q1, which focuses in $y$, was varied and the beam was observed
at the wire scanner, which was about 2.5~m downstream.  The value of
the Q2 gradient was chosen so that the beam was contained in the $x$
direction for all values of Q1.  For characterizing $x$, Q2 was
varied.

As the quadrupole strength is increased, the beam size at the
wire scanner goes through a minimum.  At the minimum, there is a
waist at approximately the wire-scanner position.  For larger quadrupole
strengths, the waist moves upstream in the beamline.

\subsection{Measurements}

Quadrupole scans were done a number of times for a variety of beam
currents for both the $x$ and $y$ directions.  The minimum beam size
at the wire scanner was near 2~mm, which was almost equal to the size
of the steering jitter.  Approximately ten quadrupole settings were
used for each scan.  Data were recorded and analyzed off line.

\subsection{Fitting to Data}

To determine the phase-space properties of the beam at the exit of the
{\sc rfq}, we needed a model that could predict the beam profile at
the wire scanner, given the beam at the {\sc rfq} exit.  We
parameterized the {\sc rfq} beam with the Courant-Snyder parameters
$\alpha$, $\beta$, and $\epsilon$ in the three directions.  We used
the simulation codes {\sc trace 3-d} and {\sc linac} as models for
computing rms beam sizes in our fitting. The {\sc trace 3-d}
code is a sigma-matrix (second moments) code that includes only linear
effects but is 3-D.  The {\sc linac} code is a particle in cell ({\sc
pic}) code that has a nonlinear $r$-$z$ space charge algorithm.

\begin{figure}[htb]
\centering
\includegraphics[angle=-90,width=0.8\columnwidth]{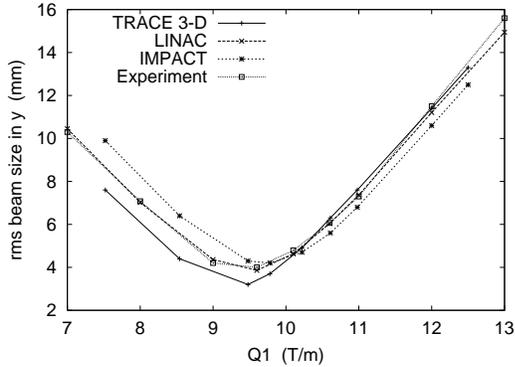}
\caption{Rms beam size at wire scanner as function of quad strength.
         All simulations used the fit to the {\sc linac} model for the
         input beam.}
\label{t1}
\end{figure}

Figure \ref{t1} shows the rms beam size in the $y$ direction as a
function of Q1 gradient.  The experimental numbers are averages from a
set of quad scan runs\cite{r4}.  The other curves are simulations
using the {\sc trace 3-d}, {\sc linac}, and {\sc impact} codes.  The
{\sc impact} code is a 3-D {\sc pic} code with nonlinear space charge.
The initial beam (at the {\sc rfq} exit) for all simulations is the
beam determined by the fit to the {\sc linac} model\cite{r4}. (This
is why there is little difference between the experimental points and
the {\sc linac} simulation.)  There are significant differences among
the codes in the predictions of the the rms beam size.
Table 1 shows emittances we obtained when fitting to the {\sc trace
3-d} and {\sc linac} models.
\begin{center}
\begin{tabular}{|l|c|c|}
\multicolumn{3}{c}{Table 1: Rms normalized emittances (mm$\cdot$mrad)}\\\hline
                              & $\epsilon_x$    & $\epsilon_y$     \\ \hline
Prediction ({\sc parmteqm})   & 0.245           & 0.244            \\
Measured ({\sc trace 3-d} fit)& 0.400           & 0.401            \\
Measured ({\sc linac} fit)    & 0.253           & 0.314            \\ \hline
\end{tabular}
\end{center}

\section{QUAD SCAN SIMULATIONS}

\subsection{Profiles at Wire Scanner}

\begin{figure}[htb]
\centering
\includegraphics[angle=-90,width=.48\columnwidth]{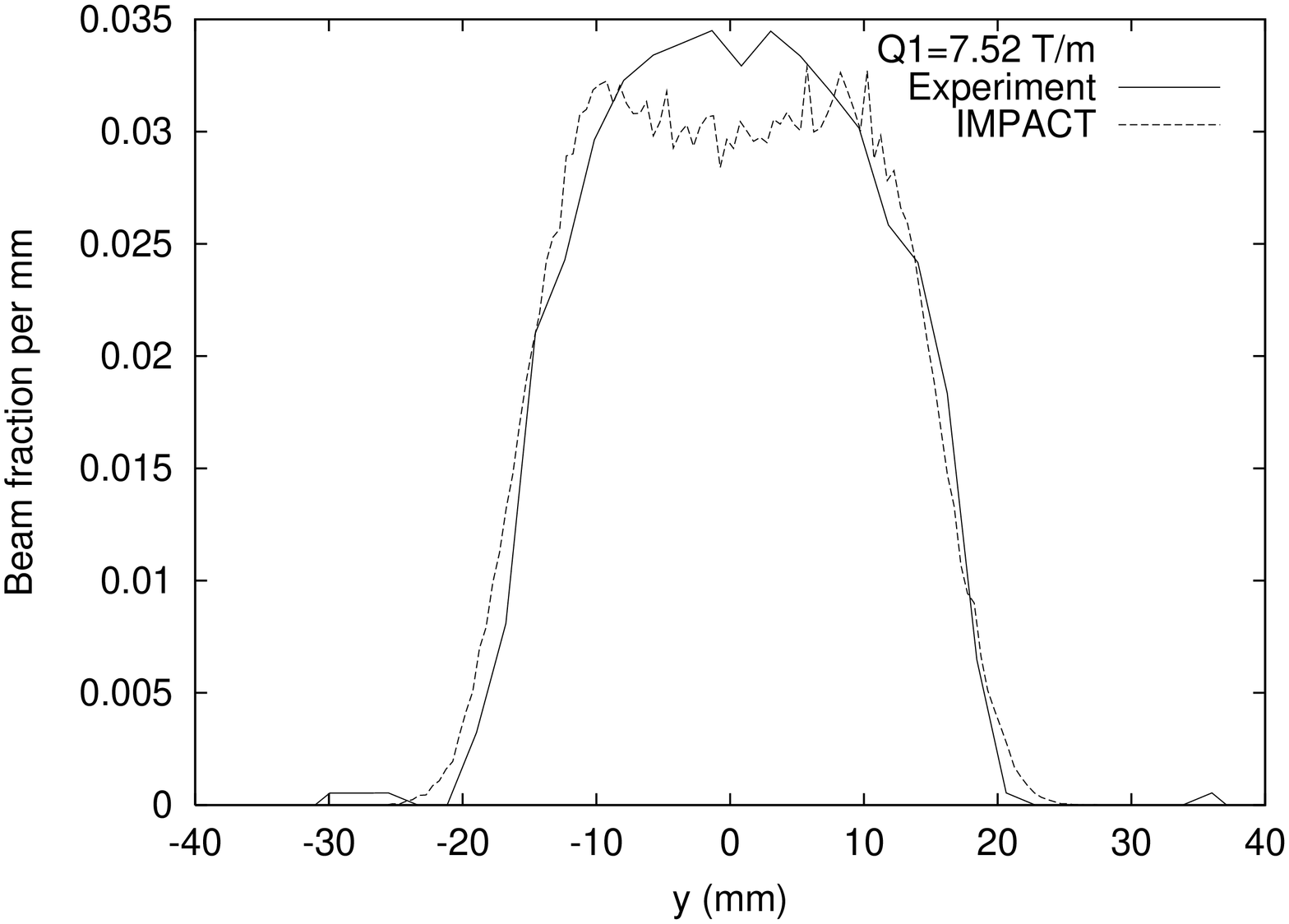}
\includegraphics[angle=-90,width=.48\columnwidth]{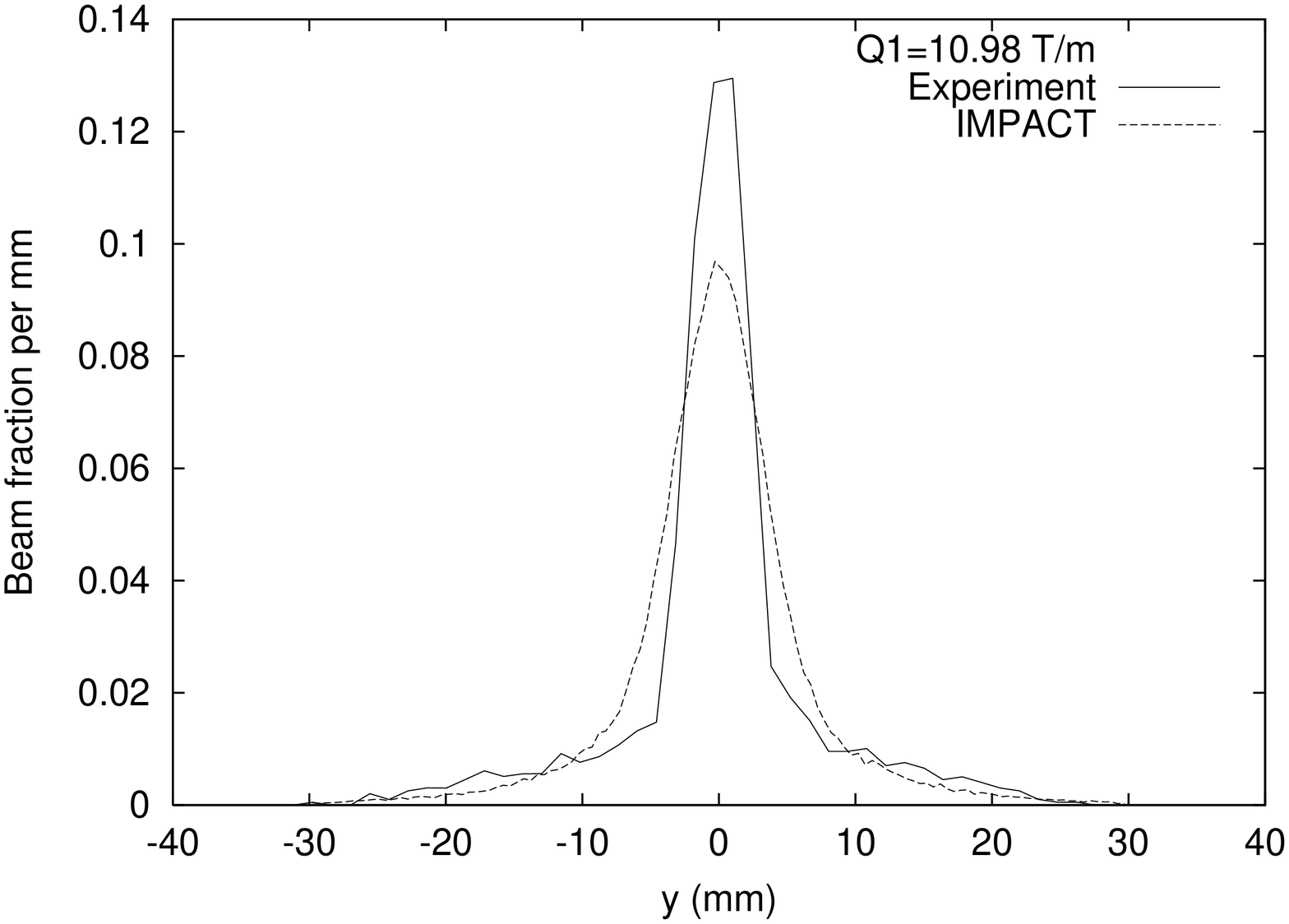}
\caption{Profile at wire scanner for $y$ scan with Q1=7.5 T/m (left)
         and Q1=11 T/m (right). Solid curve is the experimental
         measurement and the dashed curve is the {\sc impact}
         simulation using the {\sc linac}-fit beam as input.}
\label{f3}
\end{figure}

Since only the {\sc impact} code has nonlinear 3-D space charge, we
would expect that this code would be the most accurate and should be
used to fit to the data.  Both nonlinear and 3-D effects are large in
the quad scans.  However, we found that the {\sc impact} code (as well
as {\sc linac}) could not predict well the beam profile at the wire
scanner.  Figure~\ref{f3} shows the projections onto the $y$ axis for
two points of the $y$ quad scan, corresponding to a Q1 gradients of
7.52 and 11.0 T/m.  The agreement for 11 T/m, which is to the right of
the minimum of the quad scan curve, is especially poor. We see that
the experimental curve (solid) has a narrower peak, with more beam in
the tail than the {\sc impact} simulation predicts.

\begin{figure}[htb]
\centering
\includegraphics[angle=-90,width=.45\columnwidth]{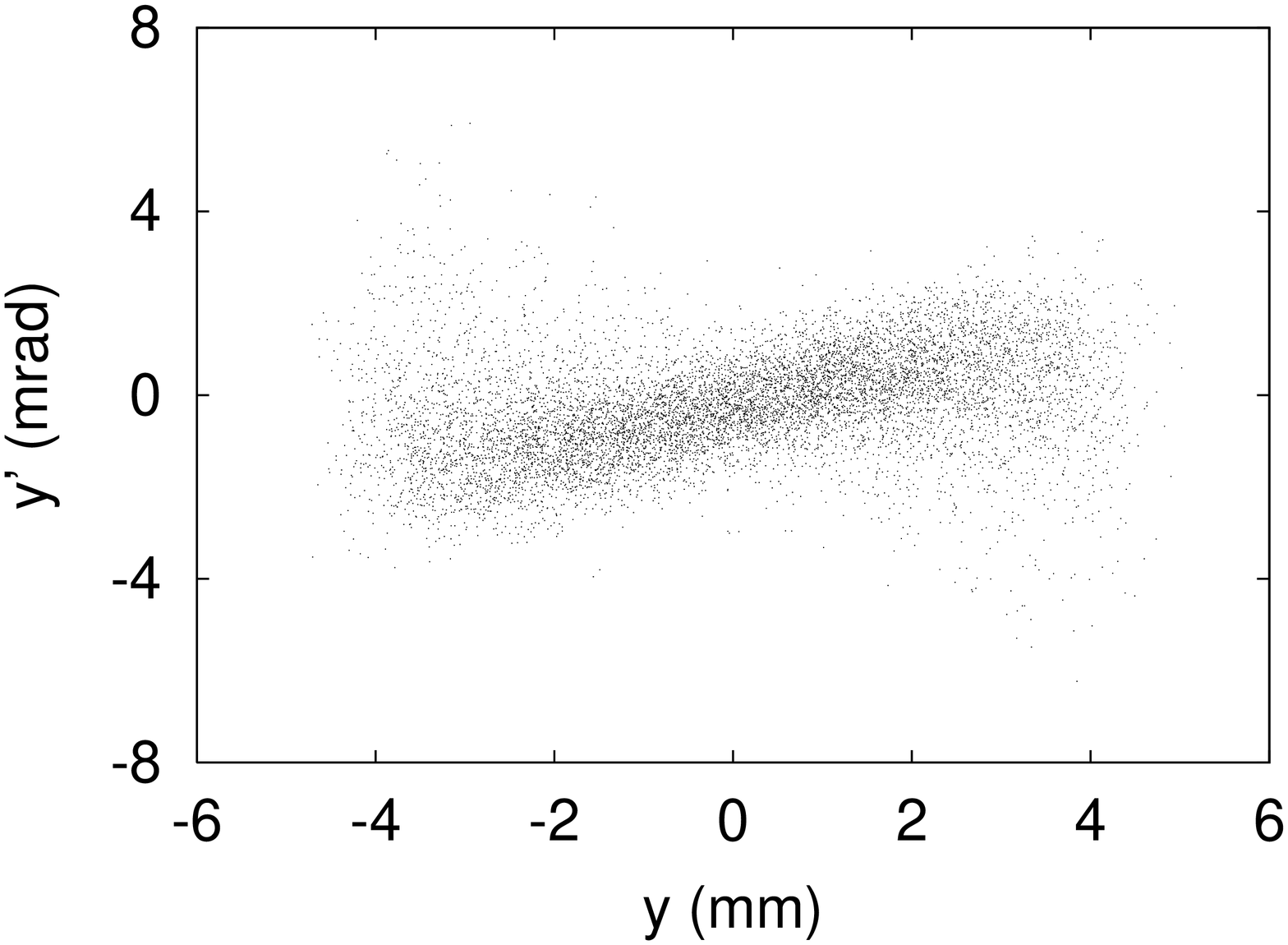}
\includegraphics[angle=-90,width=.45\columnwidth]{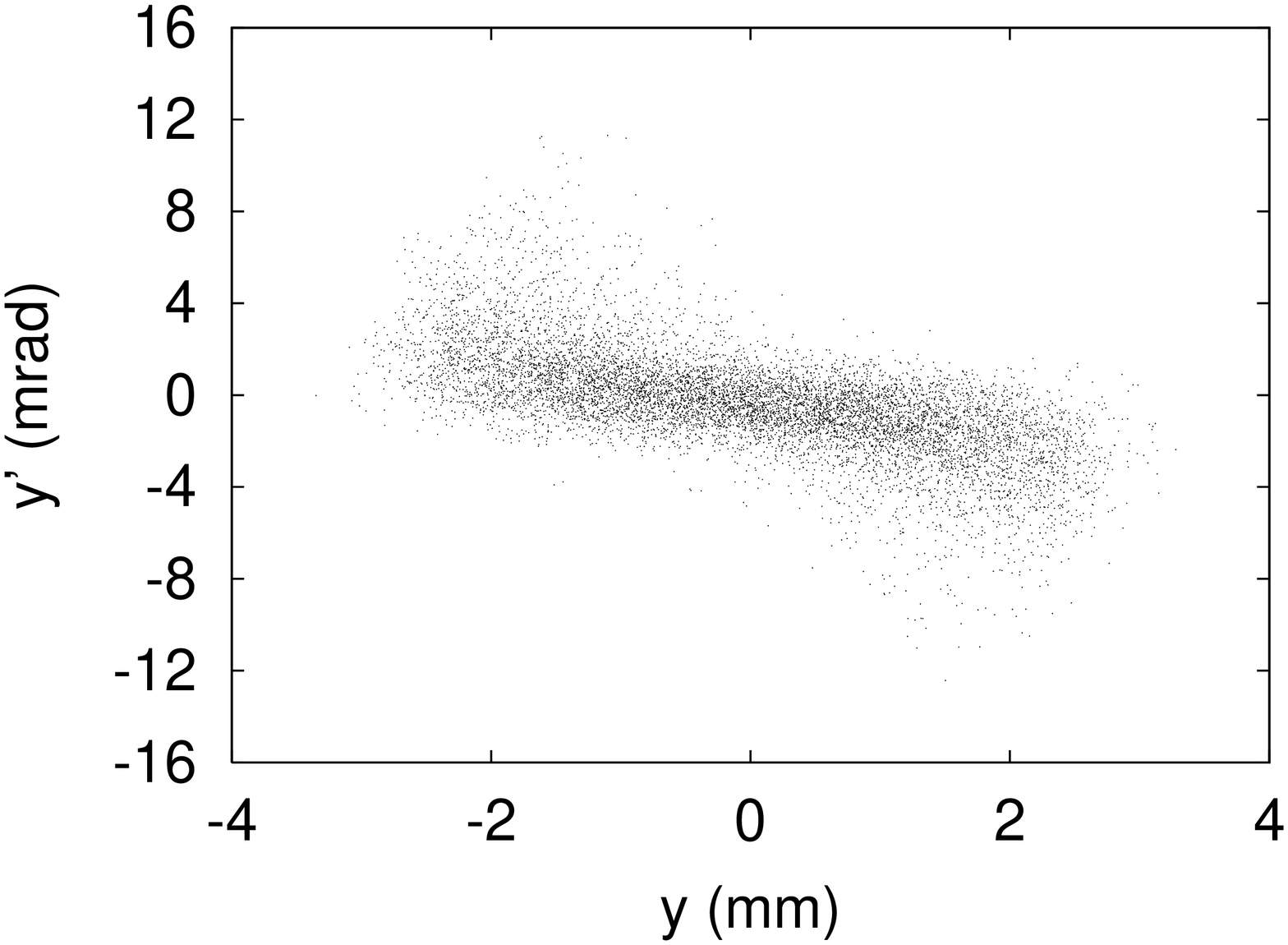}
\caption{Phase space after Q2 in $y$ direction for $y$ scan with
         Q1=9.5 T/m (left) and Q1=11 T/m (right).}
\label{f1}
\end{figure}

\begin{figure}[htb]
\centering
\includegraphics[angle=-90,width=.45\columnwidth]{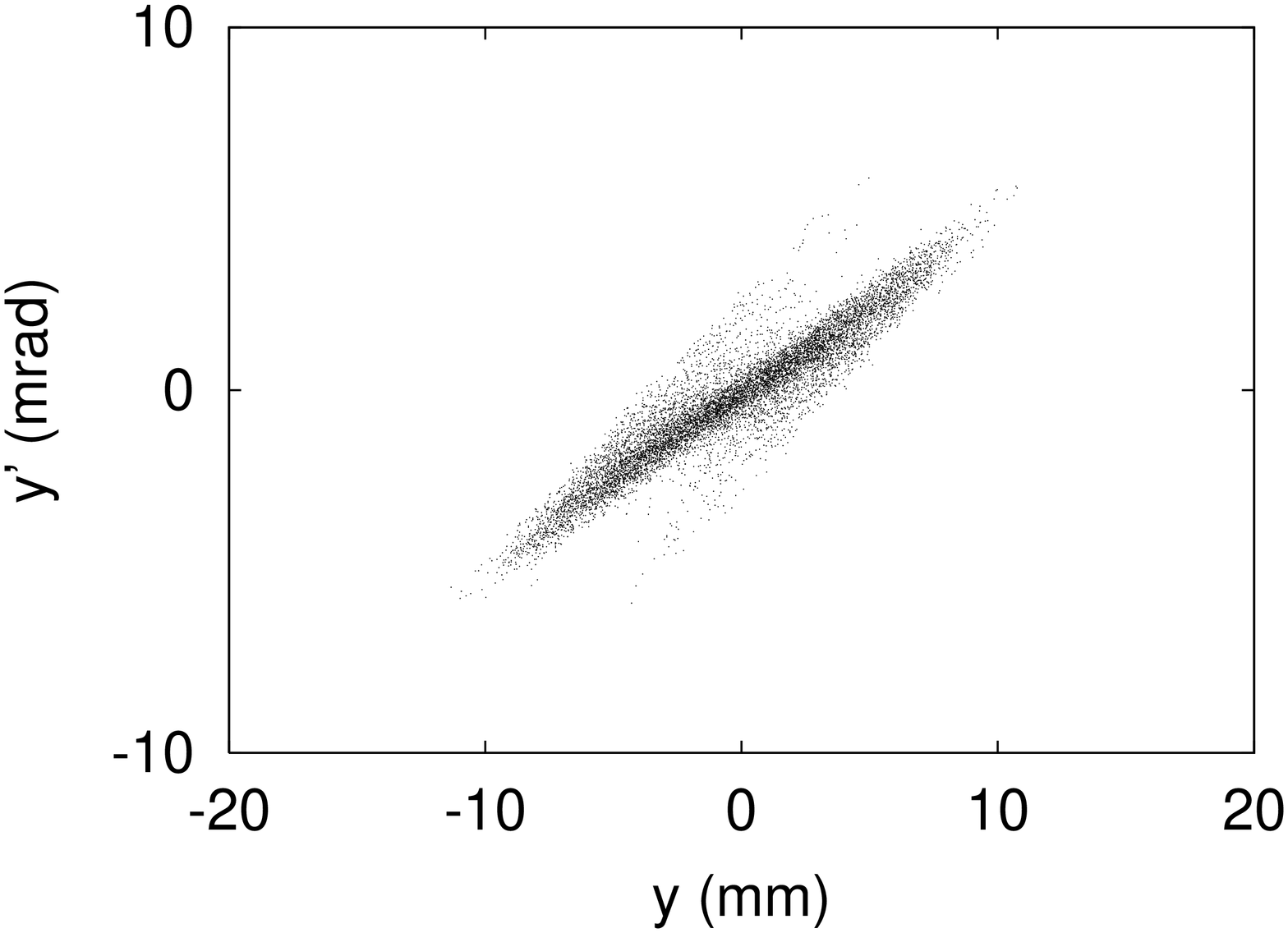}
\includegraphics[angle=-90,width=.45\columnwidth]{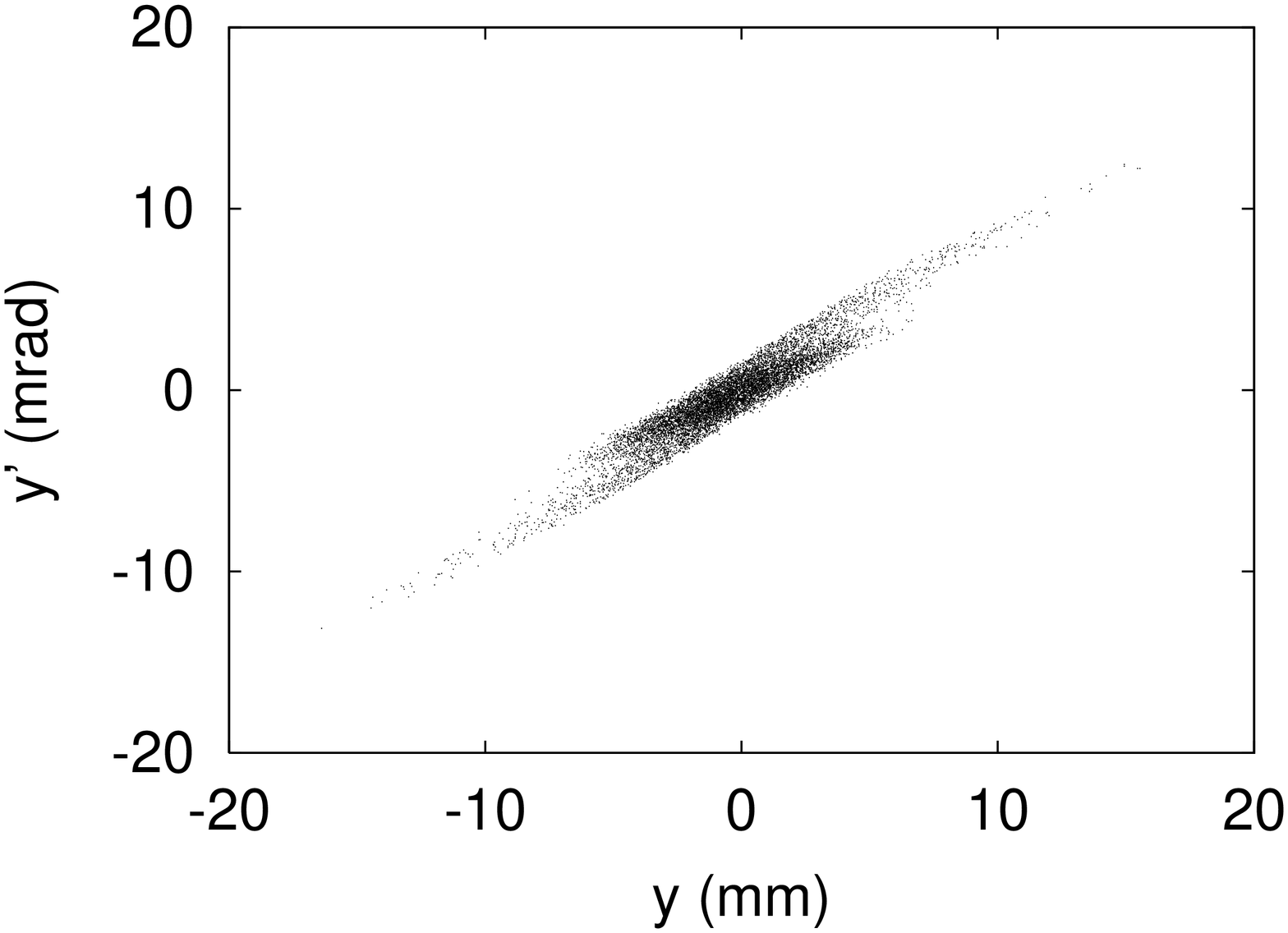}
\caption{Same as Fig. \ref{f1} but at a point just upstream of the
         wire scanner.}
\label{f2}
\end{figure}

Figure \ref{f1} shows the $y$ phase space just after Q2 for two points
in the $y$ quad scan.  After Q2, space charge has little effect and
the beam mostly just drifts to the end (there is little change in the
maximum value of $|y'|$).  The graph on the left is for a Q1 value to
the left of the quad scan minimum (9.5 T/m).  The graph at the right
shows the situation to the right of the minimum (10.9 T/m).  The
distribution in the left graph is diverging, while the one on the
right is converging.  It is this convergence that apparently leads to
the strange tails we seen in the experimental profiles at the wire
scanner.
Figure~\ref{f2} shows similar graphs a little before the wire scanner,
2.35~m downstream of the {\sc rfq}.  We see how the tails in the $y$
projection form for the case of the quad scan points to the right of
the minimum, which correspond to larger quad gradients.
While this appears to explain the narrow-peak-with-enhanced-tails seen
in the wire scans, the effect is much smaller than in the experiment.

We studied various effects looking to better reproduce the profiles
seen at the wire scanner, all with negative results.

\subsection{Code Physics}

We studied the effects of mesh sizes, boundary conditions, particle
number, and time step sizes with no significant change in results.

We investigated the possibility that there were errors associated with
using normalized variables ($p_x$) in a $z$ code, which {\sc impact}
is.  For high-eccentricity ellipses, this could be problem.  However,
transforming distributions to unnormalized coordinates, which are
appropriate to a $z$ code, did not noticeably change the results.

\subsection{Effects of Input Beam}

We used for input the beam generated by the {\sc rfq} simulation code
{\sc parmteqm}.  We also used generated beams, which were specified by
the Courant-Snyder parameters.  Using the Courant-Snyder parameters of
the {\sc parmteqm} beam yielded similar results.  Varying these
parameters in various ways did not make the beam look any closer to
the experimentally observed one.

We tried various distortions of the input beam such as enhancing the
core or tail and distorting the phase space by giving each particle a
kick in $y'$ direction proportional to $y^2$ or $y^3$.  These changes
had little effect, even for very severe distortions.  Kicks
proportional to $y^{1/3}$ were more effective.  These are more like
space-charge effects in that the distortion is larger near the origin
and smaller near the tails.  In general, we found that any structure
we put into the input beam tended to disappear because of the strong nonlinear
space-charge forces at the {\sc hebt} front end.

\subsection{Effects of Quad Errors}

Multipole errors were investigate using a version of {\sc marylie}
with 3-D space charge. We could generate tails that looked like the
experimentally observed ones, but this took multipoles that were about
500 times as large as were measured when the quadrupoles were mapped.

Quadrupole rotation studies also yielded negative results.

\subsection{Space Charge}

We investigated various currents and variations in space charge
effects along the beamline, as could be generated by neutralization or
unknown effects.

\subsection{Longitudinal Motion}

We had practically no knowledge of the beam in the longitudinal
direction except that practically all of the beam is very near the
6.7~MeV design energy.  Since the transverse beam seems to be reasonably
predicted by the {\sc rfq} simulation code, we do not expect the
longitudinal phase space to be much different from the prediction.  We
tried various longitudinal phase-space variations and none led to
profiles at the wire scanner that looked similar to the experimental ones.

\section{DISCUSSION}

In the upstream part of the {\sc hebt} the beam size profiles
($x_\mathrm{rms}$ and $y_\mathrm{rms}$ as functions of $z$) for the
quad scan tune are not much different from those of the normal {\sc
hebt} tune.  The differences occurs quite a way downstream. But here,
space charge effects are small and are unlikely to explain the
differences we see in the beam profiles at the wire scanner.  This is
a mystery that is still unresolved.

If we succeed in simulating profiles at the
wire scanners that look more like the ones seen in the measurement,
then it will be reasonable to fit the data to the 3-D {\sc impact}
simulations.  In that case, we will use all the wire-scanner data,
taking into account the detailed shape of the profile and not just the
rms value of the beam width, as we did for the {\sc trace 3-d} and
{\sc linac} fits.  While we were able to use a personal
computer to run the {\sc hpf} version of {\sc impact} for most of the
work described here, the fitting to the {\sc impact} model will have
to be done on a supercomputer.

\section{ACKNOWLEDGEMENTS}

We thank Robert Ryne and Ji Qiang for providing the {\sc impact} code
and for help associated with its use.


\end{document}